\pdfoutput=1
\documentclass[atlasdraft=false, USenglish, texlive=2023, cernpreprint, texmf, orcidlogo]{atlasdoc}
 
\usepackage[full]{atlaspackage}
\usepackage{atlasbiblatex}
 
\usepackage{atlasphysics}

\graphicspath{{logos/}{figures/}}

\AtlasTitle{Observation of an excess of di-charmonium events in the four-muon final state with the ATLAS detector}
 
\AtlasVersion{2.3}
 
\AtlasAbstract{
A search is made for potential $cc\bar{c}\bar{c}$ tetraquarks decaying into a pair of charmonium states in the four muon final state
using proton--proton collision data at $\sqrt{s}=13$~TeV, corresponding to an integrated luminosity of 140~fb$^{-1}$ recorded by the ATLAS
experiment at LHC\@. Two decay channels, $J/\psi+J/\psi \rightarrow 4\mu$ and $J/\psi+\psi(2S) \rightarrow 4\mu$, are studied. Backgrounds are estimated
based on a hybrid approach involving  Monte Carlo simulations and data-driven methods.
Statistically significant excesses with respect to backgrounds dominated by the single parton scattering are seen in the di-$J/\psi$ channel consistent with a narrow resonance at 6.9~GeV~and a
broader structure at lower mass. A statistically significant excess is also seen in the $J/\psi$+$\psi$(2S) channel. The fitted masses and decay widths of the structures are reported.
}
 
\author{The ATLAS Collaboration}
 
\AtlasRefCode{BPHY-2022-01}
 
\PreprintIdNumber{CERN-EP-2023-035}

\AtlasJournalRef{\PRL 131,151902 (2023)}
\AtlasDOI{10.1103/PhysRevLett.131.151902}

\hypersetup{pdftitle={ATLAS document},pdfauthor={The ATLAS Collaboration}}
 
\newif\ifconf
\conffalse
\begin{document}
 
\maketitle

% The next lines are included from the .//tex/introduction.tex input file
 
Beyond the conventional mesons ($q\bar{q}$) and baryons ($qqq$ or $\bar{q}\bar{q}\bar{q}$), exotic hadrons composed of four ($qq\bar{q}\bar{q}$) or
five quarks ($qqqq\bar{q}$) are also allowed under color confinement. The $X(3872)$ particle discovered by Belle in 2003 was the first tetraquark (TQ)
candidate~\cite{X3872}, and was followed by a series of further candidates designated as X, Y, and Z states~\cite{XYZ}. In 2020, LHCb observed a
narrow $X(6900)$ structure in the di-$J/\psi$ channel~\cite{LHCb_TQ}. The structure could be interpreted as a tetraquark with four charm quarks, $T_{cc\bar{c}\bar{c}}$~\cite{Iwasaki:1975pv,Chao:1980dv,Wu:2016vtq,Anwar:2017toa,Debastiani:2017msn,Liu:2019zuc,Becchi:2020uvq,Bedolla:2019zwg}.
An additional enhancement closer to the di-$J/\psi$ mass threshold was also observed in the LHCb data. Since the 6.9~\GeV~LHCb resonance is
above the $J/\psi$+$\psi$(2S) mass threshold, a structure in the $J/\psi$+$\psi$(2S) channel is also possible.
Both channels are investigated by ATLAS in a quite different phase space region from LHCb, and the new channel of $J/\psi$+$\psi$(2S) provides
more information for di-charmonium excesses. For example in some predictions, the two channels
are coupled via Pomeron exchange between the two charmonia, and $X(6900)$ is dynamically produced~\cite{ZhaoQiang}.
 
A search in the $4\mu$ final state produced through the di-$J/\psi$ and $J/\psi$+$\psi$(2S) channels is carried out, using 140 fb$^{-1}$ of LHC proton--proton ($pp$) data
collected by the ATLAS experiment at a center-of-mass energy of $\sqrt{s}=13$\nobreak\hspace{.24em}~\TeV~between the years 2015 and 2018.
Only the data where all detector systems are functional and recording high-quality data are used.
The ATLAS detector~\cite{PERF-2007-01} covers nearly the entire solid angle around the collision point\footnote{ATLAS
uses a right-handed coordinate system with its origin at the nominal interaction point (IP) and the $z$-axis along the beam pipe.
The $x$-axis points from the IP to the center of the LHC ring, and the $y$-axis points upwards.} with layered tracking detectors,
calorimeters and muon chambers. The muon and tracking systems are of particular importance in the reconstruction of charmonia.
The inner tracking detector (ID) consists of a silicon pixel detector, a silicon microstrip detector and a transition radiation tracker.
The muon spectrometer (MS) surrounds the calorimeters and consists of three large superconducting air-core toroids with eight coils
each, a system of tracking chambers, and detectors for triggering. Muons are reconstructed using information from the ID and MS systems.

% End of text imported from the .//tex/introduction.tex input file

% The next lines are included from the .//tex/simulation.tex input file
 
Background processes are estimated partly by Monte Carlo (MC) simulations and partly from data.
The main backgrounds are di-charmonium production via single parton scattering (SPS)~\cite{Qiao:2002rh,Qiao:2009kg,Lansberg1,Lansberg2,L-P-Sun,Baranov,Likhoded},
and double parton scattering (DPS)~\cite{Calucci2,Fabbro,Kom,Baranov2,Enterria,Kniehl,Borschensky}, non-prompt $J/\psi$
production from $b$-hadron decays, prompt single $J/\psi$ production and non-resonant di-muon production. \Pythia 8.244~\cite{Sjostrand:2014zea} is used
to generate SPS, DPS and non-prompt di-charmonium events. Both the color-singlet and color-octet intermediate states
are included for $J/\psi$ and $\psi$(2S). The A14~\cite{ATL-PHYS-PUB-2014-021} set of parameter values and the \texttt{NNPDF23LO}~\cite{NNPDF}
parton distribution functions (PDF)~\cite{Ball:2012cx} are used. PHOTOS 3.61~\cite{Davidson:2010ew} is applied to simulate
final state radiation in particle decays.
The remaining backgrounds which contain a single or no charmonium are modeled using the data.

% End of text imported from the .//tex/simulation.tex input file

% The next lines are included from the .//tex/selections.tex input file
 
The data sample was collected with triggers requiring either two muons with invariant mass compatible
with $J/\psi$ or $\psi$(2S) mesons (mass in the range of $[2.5, 4.3]$~\GeV), or three muons containing at least one such di-muon pair~\cite{trigger-2015,trigger-2021}.
Combinations of  triggers with different prescales~\cite{ATLAS:2020esi} depending on the run
period are used to give the largest acceptance. The trigger efficiency for the $X(6900)$ relative to the offline selection is about 72\%.
Dedicated ATLAS offline software is used to reconstruct the charmonium and $4\mu$ candidates in each event
recorded by the triggers. In each event containing at least four muons with two opposite-charge pairs, the ID tracks are fit
to a common vertex. Afterwards, each vertex of the two pairs is refit with a $J/\psi$ or $\psi$(2S)
mass constraint~\cite{ATL-PHYS-2003-031}. The resolution of the TQ mass with these mass
constraints ($m_{4\mu}$) is about 0.33\% for $X(6900)$. 
 
The \textit{loose} identification selection criteria~\cite{CERN-EP-2020-199} are required for all muon candidates.
Depending on the muon trigger thresholds and muon identification requirements, different muon momenta on the four muons are required.
Several requirements are imposed on the following variables to further suppress the background: the vertex fit quality based on $\chi^2$ per degrees of freedom $N$,
the signed distances between the primary\footnote{The primary interaction vertex is the collision vertex reconstructed excluding the 4$\mu$ candidate
tracks and with the smallest distance of closest approach in $z$ from the 4$\mu$ vertex.} and
reconstructed $4\mu$ vertices ($L_{xy}^{4\mu}$), and between the former vertices and the di-muon mass-constrained sub-vertices ($L_{xy}^{\text{di-}\mu}$).
Events with $\Delta R< 0.25$\footnote{The angular distance is defined as $\Delta R=\sqrt{\Delta\eta^2+\Delta\phi^2}$, with $\eta$ and $\phi$ being
the pseudorapidity and azimuthal angle of a particle, respectively.} between the two
reconstructed charmonia are used to study the signal, whereas events with $\Delta R\ge 0.25$ are used to
validate the shape of the $4\mu$ mass distribution for the SPS background and constrain its normalization.
The shape of the signal $4\mu$ mass distribution is not much affected by $\Delta R$ as well as muon $p_{\text{T}}$ requirements.
A summary of the kinematic requirements for the analysis regions is listed in Table~\ref{tab:selection}.
\begin{table*}[htb]
\caption{Summary of event selection requirements for different regions.}
\begin{center}
\begin{tabular}{c|c|c}
\toprule
Signal region & Control region & Non-prompt region \\ \midrule
\multicolumn{3}{c}{Di-muon or tri-muon triggers, oppositely charged muons from each charmonium,} \\
\multicolumn{3}{c}{\textit{loose} muons, $p_{\text{T}}^{1,2,3,4}>4,4,3,3$~\GeV~and $|\eta_{1,2,3,4}|<2.5$ for the four muons,} \\
\multicolumn{3}{c}{$m_{J/\psi}\in [2.94,3.25]$~\GeV, or $m_{\psi(2S)}\in [3.56,3.80]$~\GeV,} \\
\multicolumn{3}{c}{Loose vertex requirements $\chi^2_{4\mu}/N <40$ ($N=5$) and $\chi^2_{\text{di-}\mu}/N <100$ ($N=2$),} \\ \midrule
\multicolumn{2}{c|}{Vertex $\chi^2_{4\mu}/N < 3$, $L_{xy}^{4\mu} < 0.2$ mm, $|L_{xy}^{\text{di-}\mu}| < 0.3$ mm, $m_{4\mu}<11$~\GeV,} & Vertex $\chi^2_{4\mu}/N >6$, \\ \cmidrule{1-2}
$\Delta R<0.25$ between charmonia & $\Delta R\ge 0.25$ between charmonia & or $|L_{xy}^{\text{di-}\mu}| > 0.4$ mm \\
\bottomrule
\end{tabular}
\label{tab:selection}
\end{center}
\end{table*}

% End of text imported from the .//tex/selections.tex input file

% The next lines are included from the .//tex/backgrounds.tex input file
 
The SPS and DPS backgrounds contain two prompt charmonia and are modeled by MC simulations.
Because the event generator does not reproduce the data distributions well, kinematic corrections are derived from two dedicated control regions.
Since SPS (DPS) events are characterized by two charmonia which are nearby (distant) in $\eta-\phi$ space, the control region is
defined with a $4\mu$ mass sideband within $[7.5, 12]$ GeV ($[14, 24.5]$ GeV) without the $\Delta R$ requirement. The corrections are implemented by assigning event weights to MC simulations such that distributions of kinematic variables such as di-$J/\psi$ $p_{\text{T}}$, $\Delta\phi$ and $\Delta\eta$ between charmonia,
and the lowest muon $p_{\text{T}}$ match the data in the control regions. The corrections are then applied to all mass regions and the SPS modelling is validated in the $\Delta R>0.25$ control region.
 
The non-prompt background also contains two charmonia, albeit originating from $b$-hadron decays.
These typically contain a decay vertex that is displaced from the primary $pp$ interaction. This background is also modeled using
MC simulation, but normalized and validated by dedicated control regions obtained by reversing the vertex quality
requirements as shown in Table~\ref{tab:selection}.
Events from prompt single charmonium production and non-resonant di-muon production are collectively called "Others", and have
at least one charmonium candidate containing random combinations of mostly fake muons.
Fake muons are tracks, typically charged hadrons, that are misidentified as muons.
A data-driven method is used because MC simulations do not accurately estimate this kind of background.
A fake muon control region from data is used to model the Others background, which is defined by requiring that one charmonium candidate contains a track that is not reconstructed as a muon candidate, with all the other requirements kept unchanged. Events in the charmonium mass sidebands are used for both the normalization and shape corrections for Others.
 
In the di-$J/\psi$ channel, events from resonances in the $J/\psi$+$\psi$(2S) channel via $\psi$(2S)$\to J/\psi+X$, $\psi$(2S)$\to\gamma\chi_{cJ}$,
and $\chi_{cJ}\to\gamma J/\psi$, where particles other than di-$J/\psi$ are ignored, are included
as the feed-down background. The feed-down events normalization in di-$J/\psi$ ($N_{\text{fd}}$) and the fitted signal yield in $J/\psi$+$\psi$(2S) ($N$) are related by
\begin{equation}
\label{eq:eq8a}
N_{\text{fd}} = \frac{\mathcal{B}^\prime \epsilon^\prime}{\mathcal{B}\left(\psi(2S)\to\mu\mu\right) \epsilon} N,
\end{equation}
where $\epsilon$ ($\epsilon^\prime$) is the signal (feed-down) efficiency in $J/\psi$+$\psi$(2S) (di-$J/\psi$), and the
branching fraction $\mathcal{B}^\prime = \left[\mathcal{B}\left(\psi(\text{2S})\to J/\psi+X\right) + \mathcal{B}\left(\psi(\text{2S})\to\gamma\chi_{cJ}\right)\mathcal{B}(\chi_{cJ}\to\gamma J/\psi)\right]\mathcal{B}(J/\psi\to\mu\mu)$, where $X=\pi^+\pi^-, \pi^0\pi^0, \eta, \pi^0$. The systematic uncertainty on $N_{\text{fd}}$ is dominated by the uncertainty on $N$.

% End of text imported from the .//tex/backgrounds.tex input file

% The next lines are included from the .//tex/fitstrategy.tex input file
 
Unbinned maximum likelihood fits are performed to extract the signal information from data in the $4\mu$ mass spectra. The likelihood used for the fit is
\begin{equation}
\label{eq:eq6}
\mathcal{L} = \mathcal{L}_{SR}\left(\vec{\theta},\vec{\lambda}\right) \cdot \mathcal{L}_{CR}\left(\vec{\theta}\right) \cdot \prod_{j=1}^{K} G\left(\theta_j^\prime;\theta_j,\sigma_j\right),
\end{equation}
where $\mathcal{L}_{SR}$ ($\mathcal{L}_{CR}$) is the likelihood in the signal (control) region, $\vec{\lambda}$ are the parameters of
interest, $\theta_j$ are nuisance parameters (NP) which account for systematic uncertainties shared between the two regions.
Each NP has a Gaussian distribution constraint with a subsidiary measurement $\theta_j^\prime$, a mean $\theta_j$ and a width set to $\sigma_j=1$ by construction.
Only the background yields in the control regions are used in simultaneous fits with the signal regions. Background yields in the two regions are related by a transfer factor obtained from MC predictions and data-driven estimations, with systematic variations for both components.
 
In the di-$J/\psi$ channel, the feed-down normalized by Eq.~\ref{eq:eq8a} is included as an additional background,
and two fit models are considered. In model A, the signal probability density function in $\mathcal{L}_{SR}$ consists
of three interfering S-wave Breit--Wigner (BW) resonances multiplied with a phase space factor and convolved with a mass resolution function, which gives
\begin{align}
\label{eq:eq8}
f_s(x) = \left|\sum_{i=0}^{2}\frac{z_i}{m_i^2 -x^2 -i m_i \Gamma_i(x)}\right|^2 \sqrt{1-\frac{4 m_{J/\psi}^2}{x^2}} \otimes R(\theta),
\end{align}
where $m_i$ ($\Gamma_i(x)$) are the masses (widths) of resonances, $z_i$ are complex numbers representing the
relative magnitudes and phases ($z_1$ is fixed to unity with zero phase for this purpose), $\Gamma_i(x)=\Gamma_i \frac{m_i}{x}\frac{q}{q_i}$, where $q$ ($q_i$)
is the momentum of one charmonium in the rest frame of the di-charmonium system at the invariant mass equal to $x$ ($m_i$)~\cite{PDG}, and $R$ is the mass
resolution function. The $m_i$ terms are ordered by the subscripts.
In model B, two resonances are considered. The first one interferes with the SPS background, while the second is standalone. The signal+SPS
probability distribution function gives
\begin{equation}
\label{eq:eq8b}
f(x) = \left(\left|\frac{z_0}{m_0^2 - x^2 -i m_0 \Gamma_0(x)} + A(x) e^{i\phi} \right|^2 + \left|\frac{z_2}{m_2^2 -x^2 -i m_2 \Gamma_2(x)}\right|^2\right) \sqrt{ 1-\frac{4 m_{J/\psi}^2}{x^2} } \otimes R(\theta),
\end{equation}
where $A(x)$ and $\phi$ are the SPS background amplitude and phase relative to the resonance at $m_0$ ($|A(x)|^2$ reproduces the non-interfering SPS background from the MC prediction).
In this model, the control region becomes irrelevant and is excluded from the likelihood given in Eq.~\ref{eq:eq6}.
 
Models A and B are analogous to models I and II of the LHCb study~\cite{LHCb_TQ}, respectively.
However, interferences between the signal resonances are introduced in model A, which is not done in the analysis by LHCb.
The number of resonances in model A starts from one and increases to three with the fit quality gradually improving. A 4th resonance is added only for systematics, as the fit quality does not improve appreciably.
For comparison, a two-resonance model with interference, and a three-resonance model without interferences, are also tried.
It is found that when compared with model A, these models are excluded with a confidence level of more than 95\% based on toy MC studies.
 
In the $J/\psi$+$\psi$(2S) channel, two fit models are also considered. Model $\alpha$ assumes that the same interfering
resonances observed in the di-$J/\psi$ channel also decay into $J/\psi$+$\psi$(2S), in addition to a standalone fourth resonance in this channel. The signal
probability distribution function gives
\begin{equation}
\label{eq:eq9}
f_s(x) = \left(\left|\sum_{i=0}^{2}\frac{z_i}{m_i^2 - x^2 -i m_i \Gamma_i(x)}\right|^2+\left|\frac{z_3}{m_3^2 -x^2 -i m_3 \Gamma_3(x)}\right|^2\right) \sqrt{ 1-\left( \frac{m_{J/\psi}+m_{\psi(2S)}}{x} \right)^2 } \otimes R(\theta),
\end{equation}
where the parameters of the first three resonances, whose contribution appears as a structure just above the $m_{J/\psi} + m_{\psi(2S)}$ mass threshold, are fixed to the values from the fit to the di-$J/\psi$ channel. In contrast, model $\beta$ assumes a single resonance in this channel (i.e., without the $z_{0,1,2}$ terms in Eq.~\ref{eq:eq9}).

% End of text imported from the .//tex/fitstrategy.tex input file

% The next lines are included from the .//tex/systematics.tex input file
 
The systematic uncertainties are classified into those affecting exclusively normalizations, and those affecting the mass spectrum shape as well.
Only the latter are relevant, since the signal and background normalizations are freely floating parameters.
The systematic uncertainties in $m_{4\mu}$, with and without the muon momentum calibration corrections, are treated as resolution uncertainties.
Because the resolution of the $m_{4\mu}$ is mass dependent and a constant mass is used in the nominal fit, resolutions in different mass ranges are
treated as systematic uncertainties. A shape uncertainty is assigned to account for bin-to-bin fluctuations from the limited MC sample size for backgrounds.
In the SPS background, a \Pythia model parameter uncertainty from \texttt{pT0timesMPI}~\cite{Sjostrand:2014zea}, which controls the suppression of the soft double charmonia
production, is assigned, and its nominal value is tuned to data in the SPS control region.
A shape uncertainty in the background due to residual di-charmonium $p_{\text{T}}$ mismodeling is applied.
Based on toy MC studies, biases from the fit in the resonance parameters are also considered as systematic uncertainties.
The P and D-wave BW functions are substituted for the S-wave for resonances away from the threshold
to estimate systematic uncertainties due to different orbital angular momentum
assumptions\footnote{The first resonance at the threshold is always assumed to be S-wave, as the data has no constraining power for its width when $L=1,2$.}.
Systematic shape variations in the $X(6900)$ in the di-$J/\psi$ channel, and in the second resonance for the $J/\psi$+$\psi$(2S) channel due to the $\Delta R$ and muon $p_{\text{T}}$
requirements are considered as well.
In the di-$J/\psi$ channel, a 4th resonance around 7.2~\GeV~(hinted by the LHCb analysis) is added to the fit, and the feed-down
background normalizations are varied according to the uncertainties in $J/\psi$+$\psi$(2S). The transfer factor uncertainty is dominated by the
SPS model parameter, so it is not treated as a separate NP\@.
In the $J/\psi$+$\psi$(2S) channel, the uncertainty in a transfer factor between the signal and control regions,
and a shape uncertainty derived from the non-prompt region due to Others (shape inconsistency), are included.
Interference between the 4th resonance and the other ones are included in systematic uncertainties. In model $\alpha$, systematic uncertainties on the lower resonance shape from the di-$J/\psi$ channel model A fit are also included.
Other systematic uncertainties such as the parton PDF and Pythia parameters affect signal and background normalizations only, and are not incorporated in the fits.

% End of text imported from the .//tex/systematics.tex input file

% The next lines are included from the .//tex/results.tex input file
 
\begin{figure}[htbp]
\centering
\includegraphics[width=0.40\textwidth]{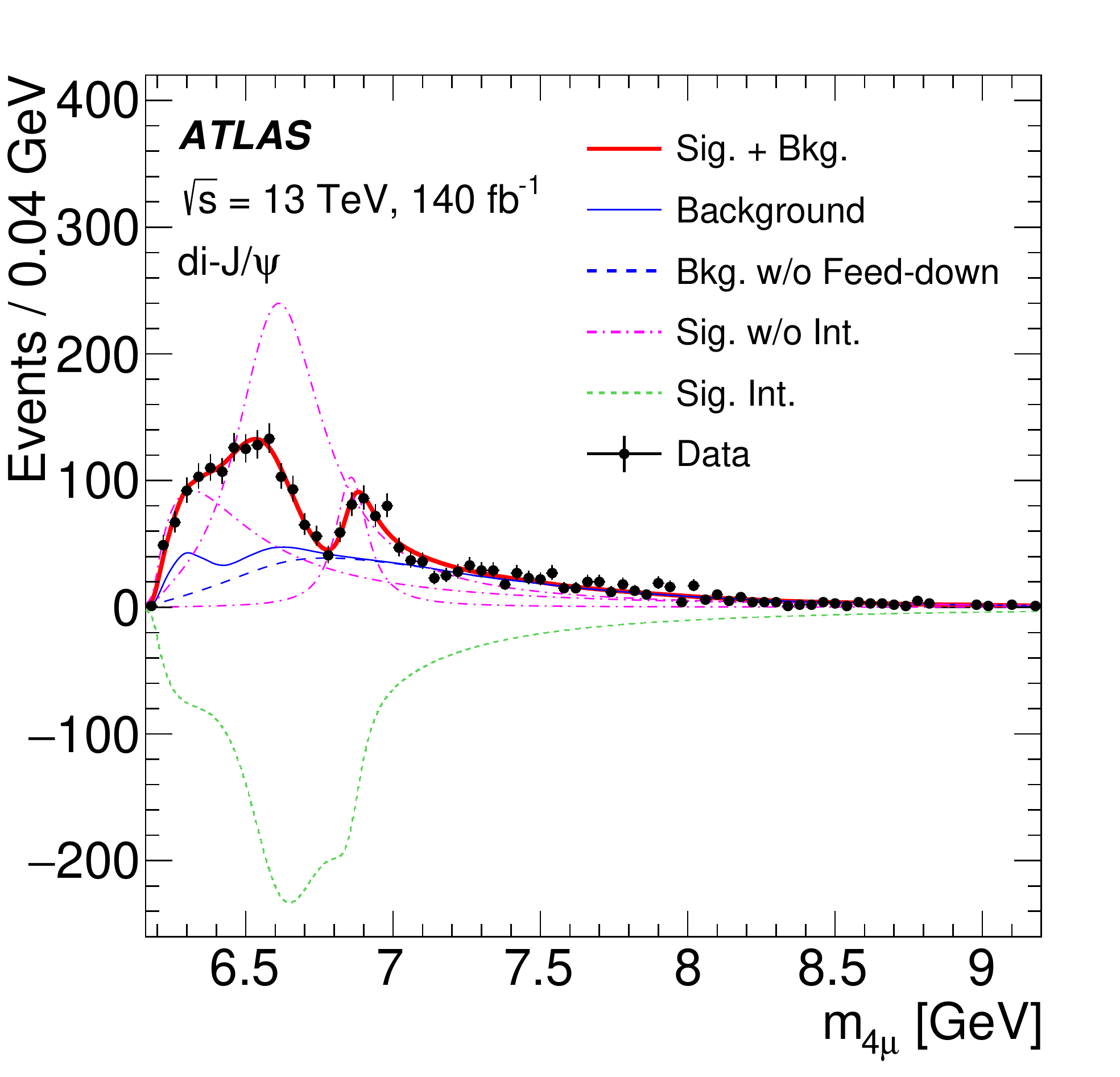}
\put(-50, 40){\textbf{(a)}}
\includegraphics[width=0.40\textwidth]{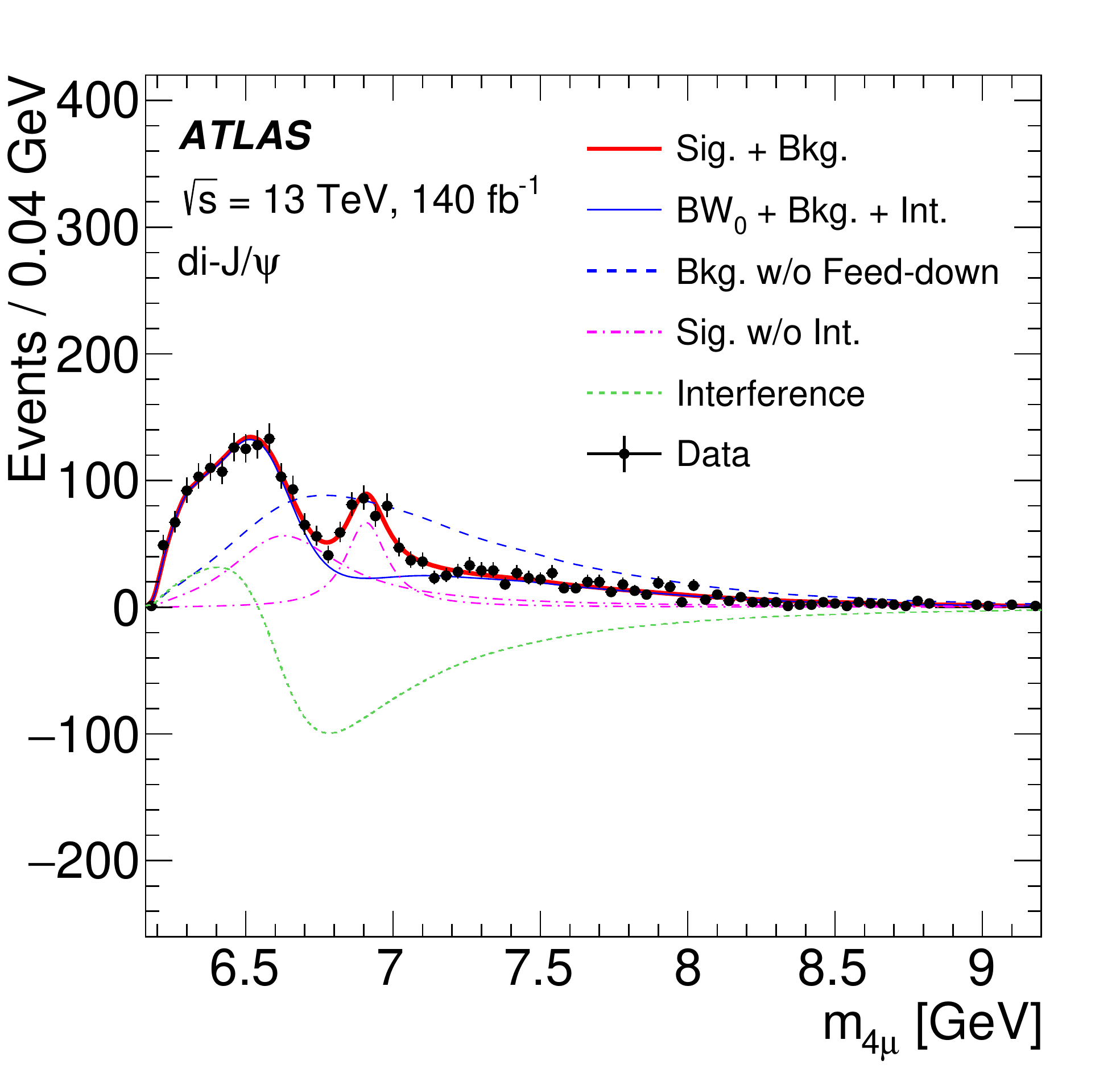}
\put(-50, 40){\textbf{(b)}}\\
\includegraphics[width=0.40\textwidth]{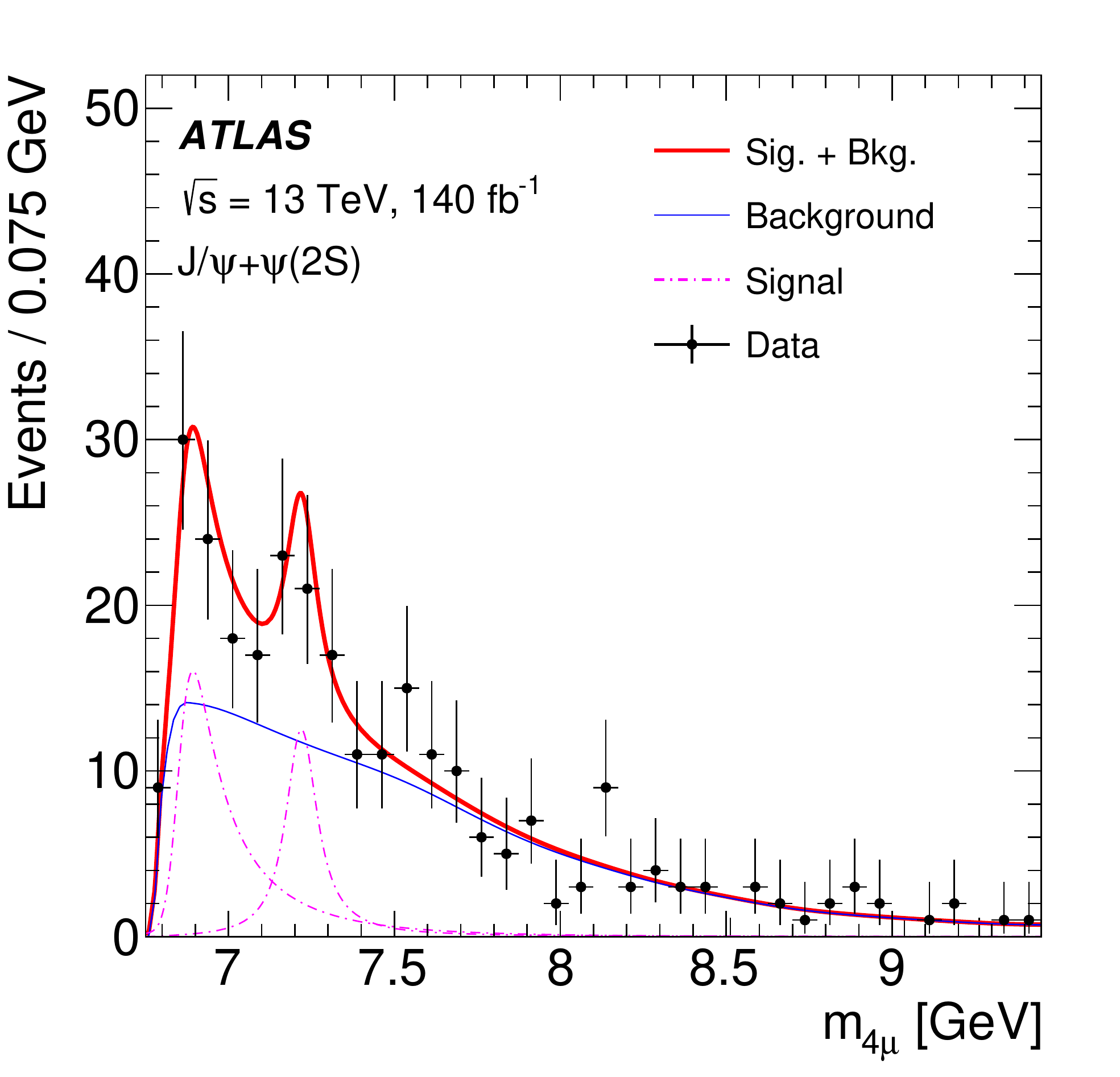}
\put(-50, 60){\textbf{(c)}}
\includegraphics[width=0.40\textwidth]{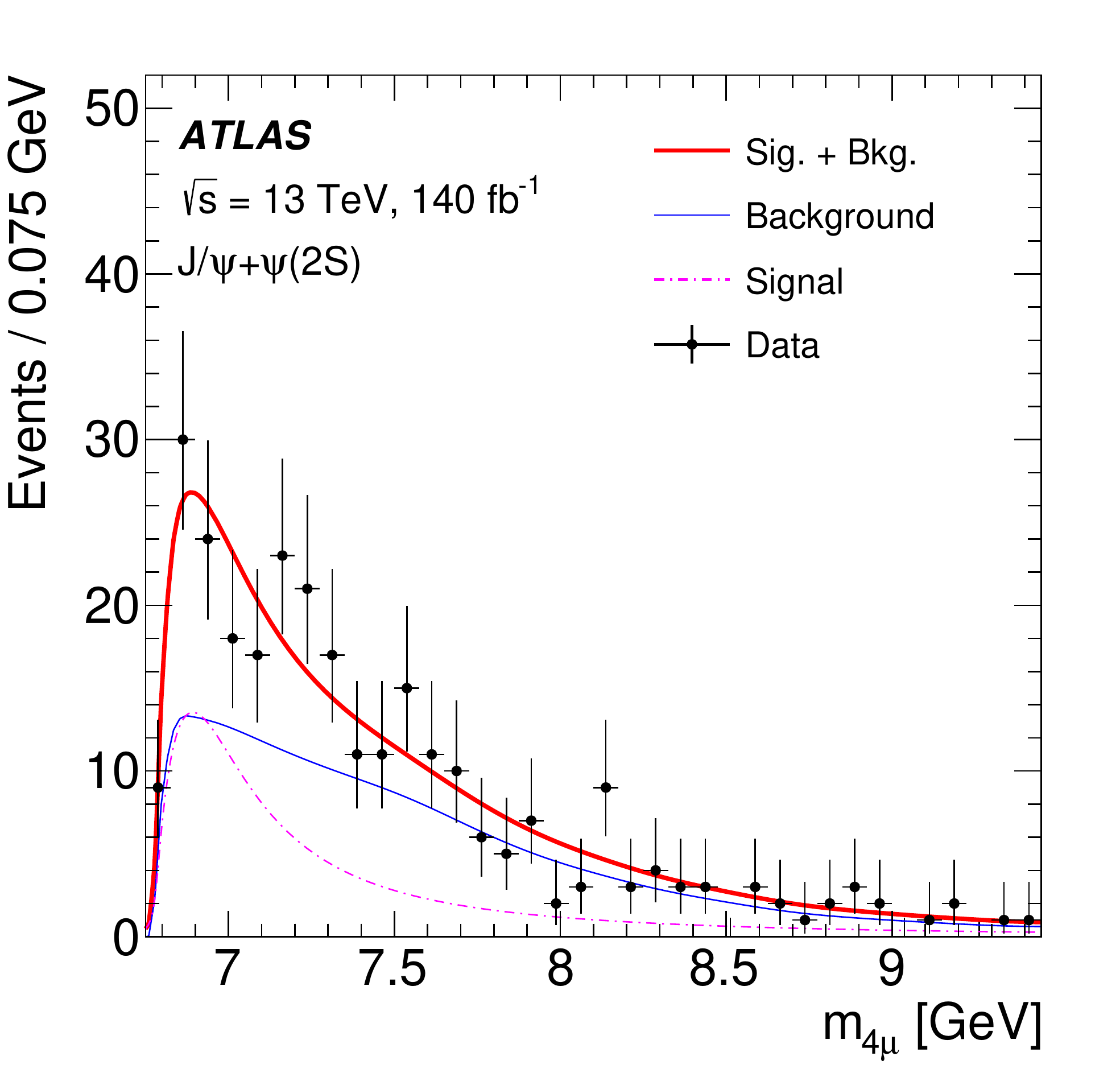}
\put(-50, 60){\textbf{(d)}}
\caption{The fit to the mass spectra in the signal regions in the di-$J/\psi$ (a,b) and $J/\psi$+$\psi$(2S) (c,d) channels. Fit results for models A (a), B (b), $\alpha$ (c) and $\beta$ (d) are shown. The purple dash-dotted lines represent the components of individual resonances, and the green short dashed ones represent the interferences among them. }
\label{fig:unb_mass_fit}
\end{figure}

The $4\mu$ mass spectra  fit to data in the two channels are shown in Figure \ref{fig:unb_mass_fit}.
The fitted masses and widths of resonances are given in Table \ref{tab:mass_width}. 
Both the significance of all resonances, and the one for $X(6900)$ alone, far exceed 5$\sigma$\footnote{The asymptotic formula based on the profile likelihood ratio, $Z = \sqrt{2\ln \frac{L(\hat{s},\hat{\theta})}{L(0,\hat{\hat{\theta}})}}$, is used to calculate the overall significance, where $s$ is the signal yield and $\theta$ are NPs~\cite{Cowan:2010js}. Similarly for $X(6900)$ alone, $Z = \sqrt{2\ln \frac{L(\hat{z}_3,\hat{\theta})}{L(0,\hat{\hat{\theta}})}}$ is used. In the calculations, the signal shape parameters are all fixed to their best-fit values.}. The mass of the third resonance, $m_2$, is consistent with the LHCb mass. Although both the models A and B describe the data well, the broad structure at the lower mass could result from other physical effects, such as the feed-down from higher di-charmonium resonances, e.g., $T_{cc\bar{c}\bar{c}}\to \chi_{cJ}\chi_{cJ'} \to J/\psi J/\psi \gamma\gamma$ where the soft photons are not reconstructed. In the $J/\psi$+$\psi$(2S) channel, 
the signal significance with signal shape parameters of model $\alpha$ ($\beta$) fixed to their best-fit values is 4.7$\sigma$ (4.3$\sigma$). In the fit with model $\alpha$, the significance of the second resonance alone is found to be 3.0$\sigma$.
\begin{table}[htbp]
\caption{ The fitted masses and natural widths (in \GeV), and relative uncertainties of signal yields ($\Delta s/s$) in the di-$J/\psi$ and $J/\psi$+$\psi$(2S) channels. The results of both the models are given in each channel. The first uncertainties are statistical while the second ones are systematic. }
\label{tab:mass_width}
\centering
\renewcommand{\arraystretch}{1.2}
\begin{tabular}{ccc}
\toprule
di-$J/\psi$ & model A & model B \\ \midrule
$m_0$ & $6.41\pm 0.08_{-0.03}^{+0.08}$ & $6.65\pm 0.02_{-0.02}^{+0.03}$ \\
$\Gamma_0$ & $0.59\pm 0.35_{-0.20}^{+0.12}$ & $0.44\pm 0.05_{-0.05}^{+0.06}$ \\
$m_1$ & $6.63\pm 0.05_{-0.01}^{+0.08}$ & \multirow{2}{*}{---} \\
$\Gamma_1$ & $0.35\pm 0.11_{-0.04}^{+0.11}$&  \\
$m_2$ & $6.86\pm 0.03_{-0.02}^{+0.01}$ & $6.91\pm 0.01\pm 0.01$ \\
$\Gamma_2$ & $0.11\pm 0.05_{-0.01}^{+0.02}$ & $0.15\pm 0.03\pm 0.01$ \\
$\Delta s/s$ & $\pm5.1\%_{-8.9\%}^{+8.1\%}$ & --- \\
\bottomrule
$J/\psi$+$\psi$(2S) & model $\alpha$ & model $\beta$ \\ \midrule
$m_3$ & $7.22\pm 0.03_{-0.04}^{+0.01}$ & $6.96\pm 0.05\pm 0.03$ \\
$\Gamma_3$ & $0.09\pm 0.06_{-0.05}^{+0.06}$ & $0.51\pm 0.17_{-0.10}^{+0.11}$ \\
$\Delta s/s$ & $\pm21\%_{-15\%}^{+25\%}$ & $\pm20\%\pm12\%$ \\
\bottomrule
\end{tabular}
\end{table}

% End of text imported from the .//tex/results.tex input file

\FloatBarrier

% The next lines are included from the .//tex/summary.tex input file
 
In conclusion, the results of a search for potential $cc\bar{c}\bar{c}$ tetraquarks decaying into a pair of $J/\psi$ charmonium states, or into a $J/\psi$ and $\psi$(2S),
in the $4\mu$ final state are presented based on $pp$ collisions data collected by the ATLAS experiment at $\sqrt{s}=13$~\TeV~corresponding to an integrated luminosity of 140 fb$^{-1}$.
A significant excess of events (far exceeding 5$\sigma$) in data above the expected background is observed in the di-$J/\psi$ channel. Analogous to LHCb observations,
a broad structure at lower mass and a resonance around 6.9~\GeV~are observed. A three-resonance model with interferences, or a model with the lower broad
structure interfering with the SPS background,  describes the excess better than models with fewer interfering resonances or with no interferences. 
In the $J/\psi$+$\psi$(2S) channel, a 4.7$\sigma$ excess of events is observed when considering a model involving two resonances, one of which is near the 6.9~\GeV~threshold.
In both channels, details of the lower-mass structure cannot be discerned directly from the data, and other interpretations
(e.g. multiple non-interfering resonances, reflection effects and threshold enhancements) cannot be excluded. More data are required to better characterize the excesses observed in both channels.

% End of text imported from the .//tex/summary.tex input file

% The next lines are included from the .//acknowledgements/Acknowledgements.tex input file

We thank CERN for the very successful operation of the LHC, as well as the
support staff from our institutions without whom ATLAS could not be
operated efficiently.
 
We acknowledge the support of
ANPCyT, Argentina;
YerPhI, Armenia;
ARC, Australia;
BMWFW and FWF, Austria;
ANAS, Azerbaijan;
CNPq and FAPESP, Brazil;
NSERC, NRC and CFI, Canada;
CERN;
ANID, Chile;
CAS, MOST and NSFC, China;
Minciencias, Colombia;
MEYS CR, Czech Republic;
DNRF and DNSRC, Denmark;
IN2P3-CNRS and CEA-DRF/IRFU, France;
SRNSFG, Georgia;
BMBF, HGF and MPG, Germany;
GSRI, Greece;
RGC and Hong Kong SAR, China;
ISF and Benoziyo Center, Israel;
INFN, Italy;
MEXT and JSPS, Japan;
CNRST, Morocco;
NWO, Netherlands;
RCN, Norway;
MEiN, Poland;
FCT, Portugal;
MNE/IFA, Romania;
MESTD, Serbia;
MSSR, Slovakia;
ARRS and MIZ\v{S}, Slovenia;
DSI/NRF, South Africa;
MICINN, Spain;
SRC and Wallenberg Foundation, Sweden;
SERI, SNSF and Cantons of Bern and Geneva, Switzerland;
MOST, Taiwan;
TENMAK, T\"urkiye;
STFC, United Kingdom;
DOE and NSF, United States of America.
In addition, individual groups and members have received support from
BCKDF, CANARIE, Compute Canada and CRC, Canada;
PRIMUS 21/SCI/017 and UNCE SCI/013, Czech Republic;
COST, ERC, ERDF, Horizon 2020 and Marie Sk{\l}odowska-Curie Actions, European Union;
Investissements d'Avenir Labex, Investissements d'Avenir Idex and ANR, France;
DFG and AvH Foundation, Germany;
Herakleitos, Thales and Aristeia programmes co-financed by EU-ESF and the Greek NSRF, Greece;
BSF-NSF and MINERVA, Israel;
Norwegian Financial Mechanism 2014-2021, Norway;
NCN and NAWA, Poland;
La Caixa Banking Foundation, CERCA Programme Generalitat de Catalunya and PROMETEO and GenT Programmes Generalitat Valenciana, Spain;
G\"{o}ran Gustafssons Stiftelse, Sweden;
The Royal Society and Leverhulme Trust, United Kingdom.
 
The crucial computing support from all WLCG partners is acknowledged gratefully, in particular from CERN, the ATLAS Tier-1 facilities at TRIUMF (Canada), NDGF (Denmark, Norway, Sweden), CC-IN2P3 (France), KIT/GridKA (Germany), INFN-CNAF (Italy), NL-T1 (Netherlands), PIC (Spain), ASGC (Taiwan), RAL (UK) and BNL (USA), the Tier-2 facilities worldwide and large non-WLCG resource providers. Major contributors of computing resources are listed in Ref.~\cite{ATL-SOFT-PUB-2021-003}.

% End of text imported from the .//acknowledgements/Acknowledgements.tex input file

\clearpage
% The next lines are included from the .//tex/appendix.tex input file
\part*{Appendix}
\addcontentsline{toc}{part}{Appendix}

Signal events are simulated with the event generator JHU~\cite{JHU} and \CTEQ{6L1} PDF~\cite{Pumplin:2002vw},
or with \Pythia and the \NNPDF{23LO} PDF\@. Feed-down backgrounds from the $J/\psi$+$\psi$(2S) channel to di-$J/\psi$ are included.
A natural width of 100~\MeV~is assumed for all the resonances with no interference between them.
An extensive software suite~\cite{ATL-SOFT-PUB-2021-001} is used in data simulation~\cite{SOFT-2010-01}, in the reconstruction and
analysis of real and simulated data, in detector operations, and in the trigger and data acquisition systems of the experiment.
The MC simulated events are weighted to reproduce the same number of $pp$ interactions per bunch crossing (pileup) and trigger conditions as occur in data.
 
\begin{figure}[htbp]
\centering
\includegraphics[width=0.33\textwidth]{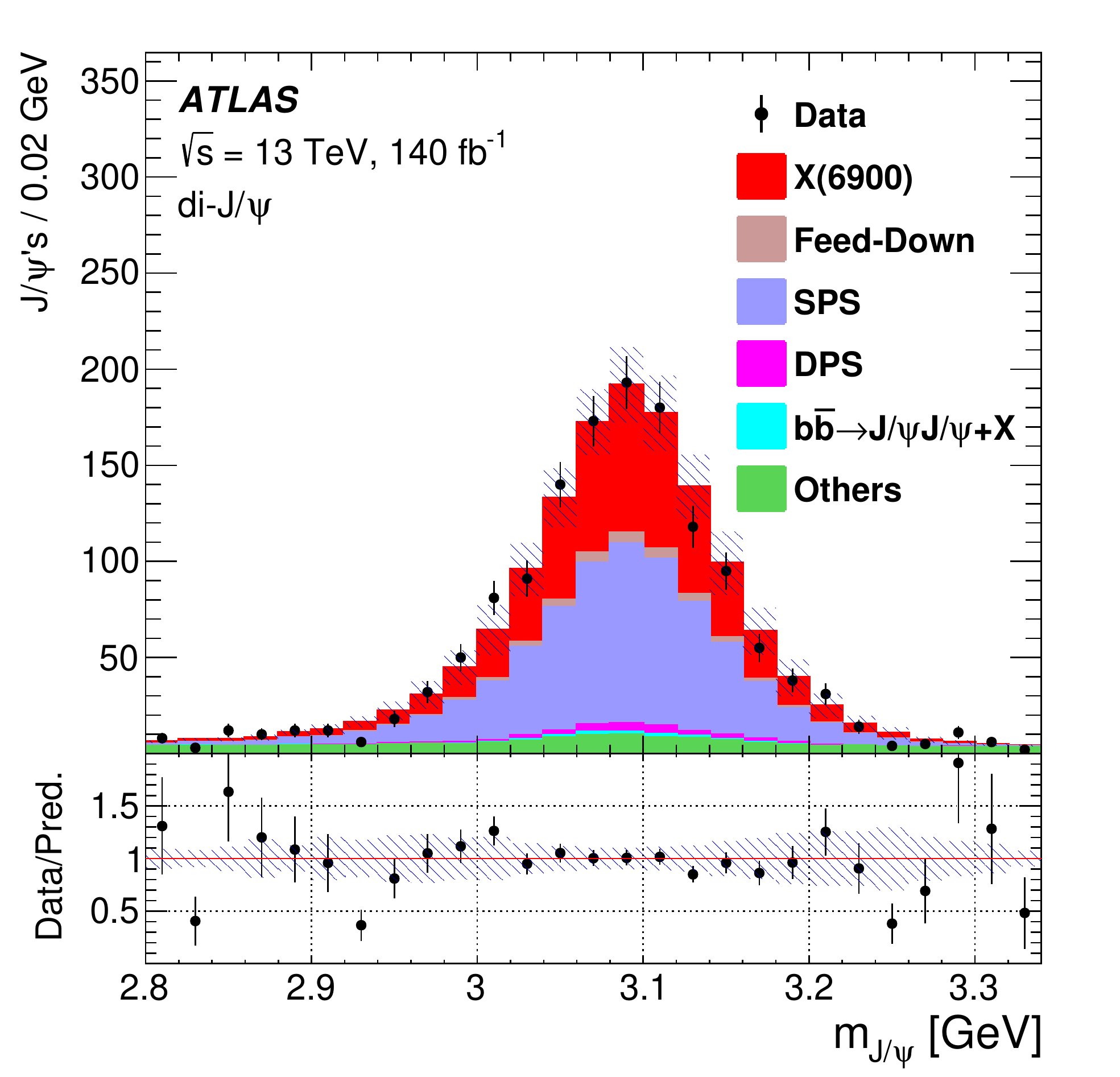}
\put(-120, 60){\textbf{(a)}}
\includegraphics[width=0.33\textwidth]{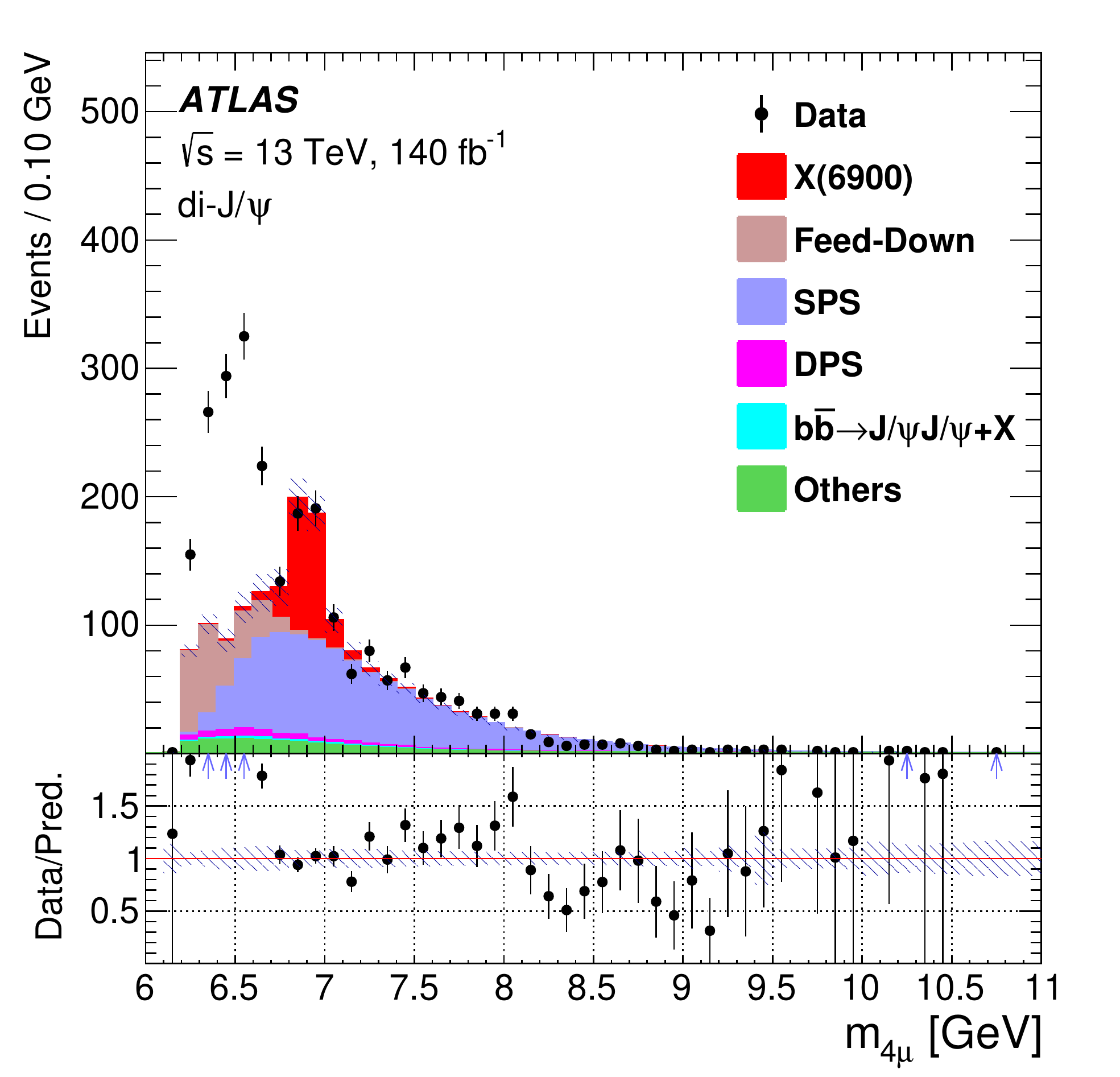}
\put(-45, 60){\textbf{(b)}}
\includegraphics[width=0.33\textwidth]{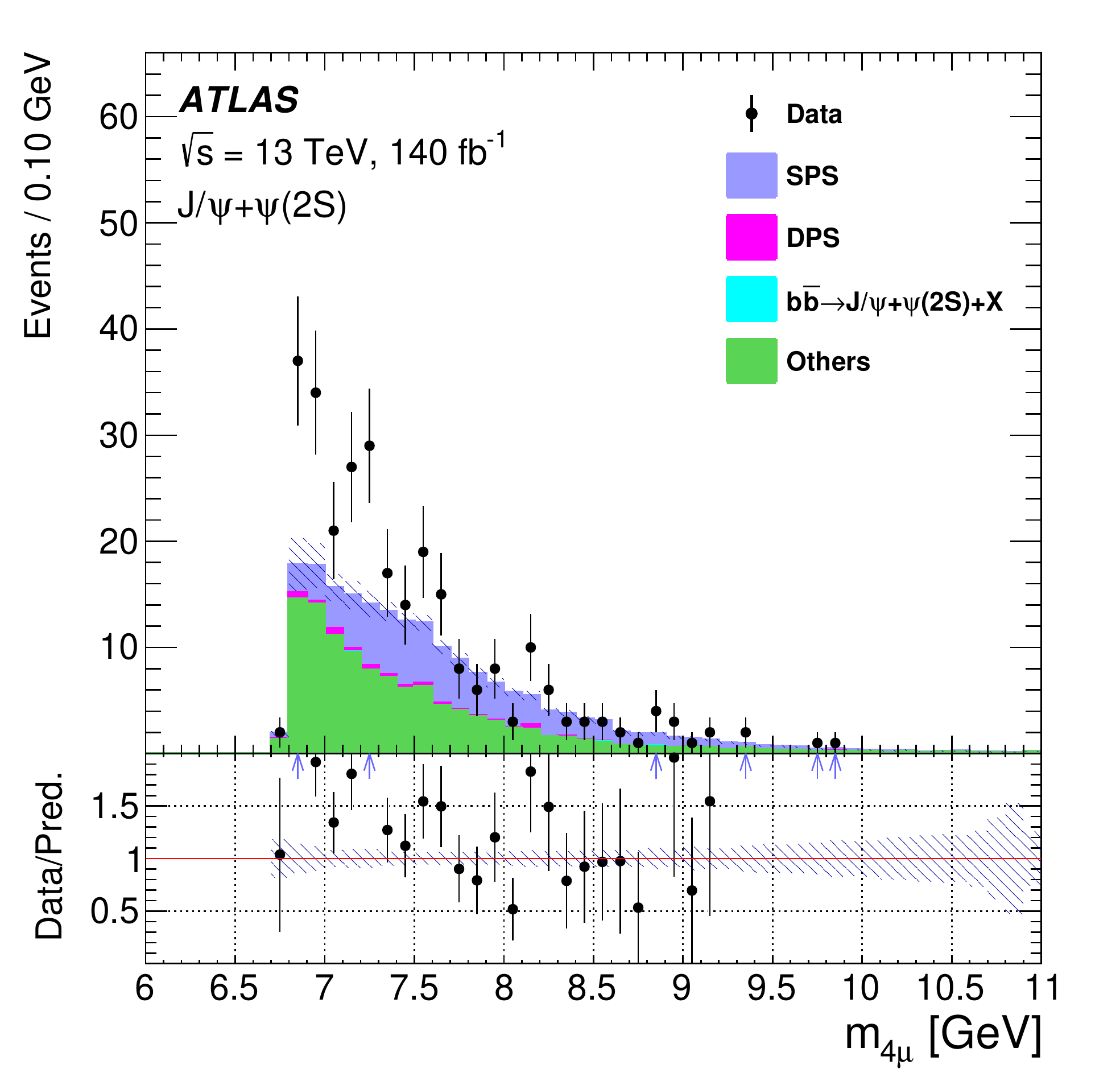}
\put(-45, 60){\textbf{(c)}}
\caption{ The $J/\psi$ mass spectrum with 6.7~\GeV$<m_{4\mu}<$7.1~\GeV~(a) and the $4\mu$ mass spectrum (b) in the signal region
in the di-$J/\psi$ channel, and the similar mass spectrum in the $J/\psi$+$\psi$(2S) channel (c). The signal from the $X(6900)$ is scaled to match
data around 6.9~\GeV. The bars and shaded areas represent uncertainties of data and predictions in each bin, respectively.
The arrows in the lower panel indicate that the ratio of data to prediction is out of range in that bin. }
\label{fig:unb_SR}
\end{figure}

\begin{figure}[htbp]
\centering
\includegraphics[width=0.35\textwidth]{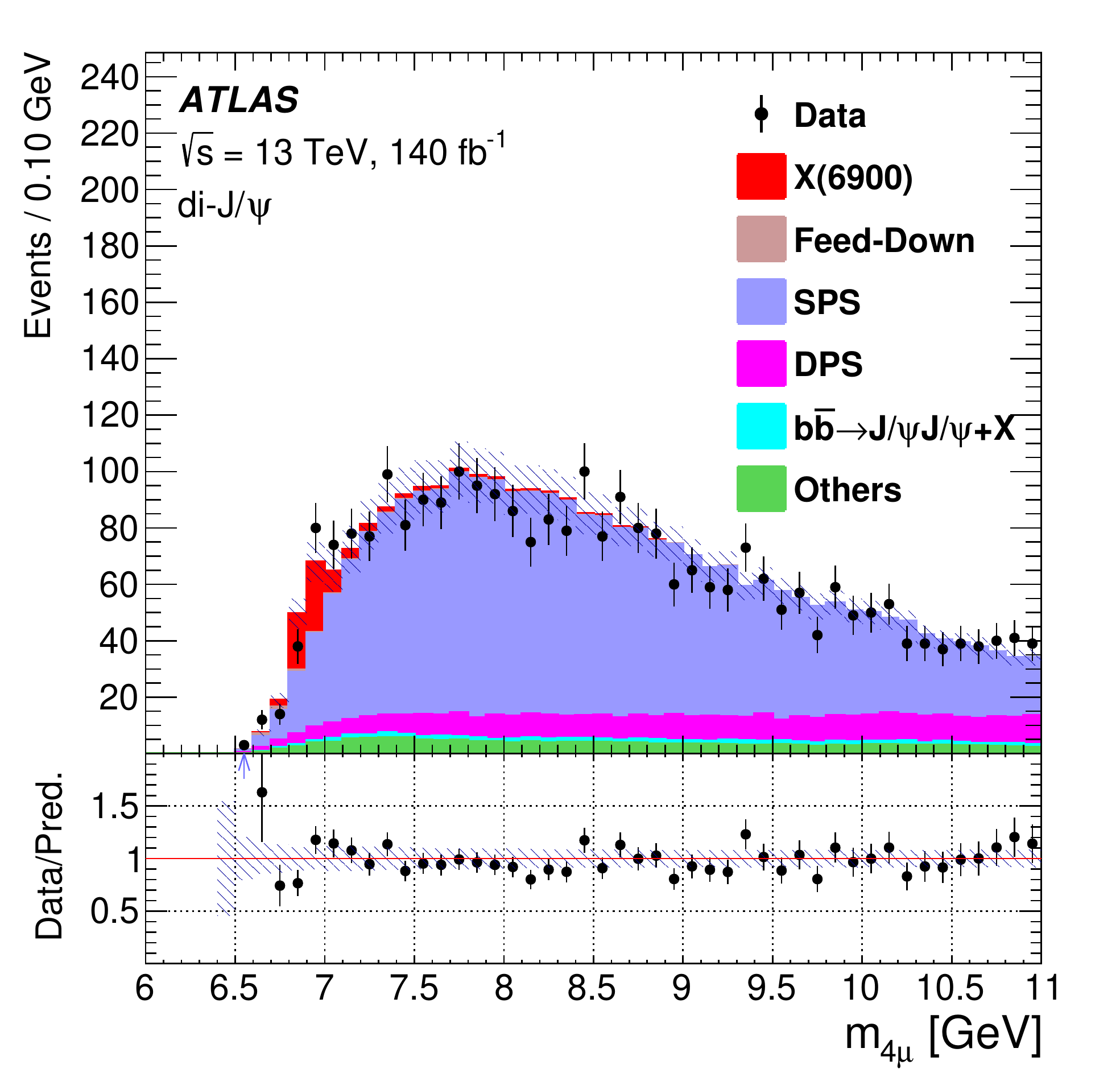}
\put(-125, 105){\textbf{(a)}}
\includegraphics[width=0.35\textwidth]{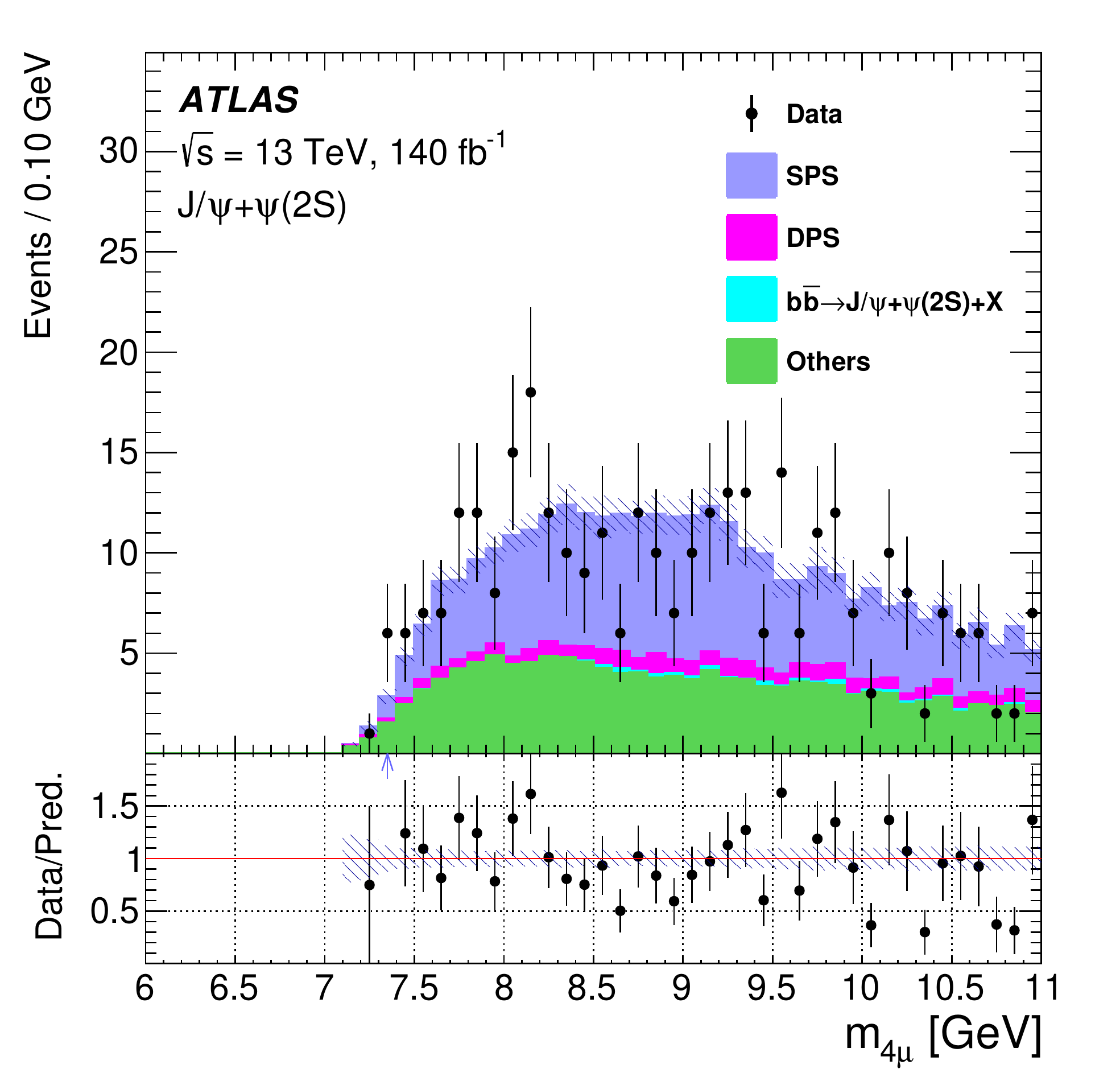}
\put(-125, 105){\textbf{(b)}}
\caption{The $4\mu$ mass spectra in the control regions with $\Delta R\ge 0.25$ in the di-$J/\psi$ (a) and $J/\psi$+$\psi$(2S) (b) channels.
The bars and shaded areas represent uncertainties of data and predictions in each bin, respectively.
The arrows in the lower panel indicate that the ratio of data to prediction is out of range in that bin. }
\label{fig:unb_CR}
\end{figure}

\begin{figure}[htbp]
\centering
\includegraphics[width=0.33\textwidth]{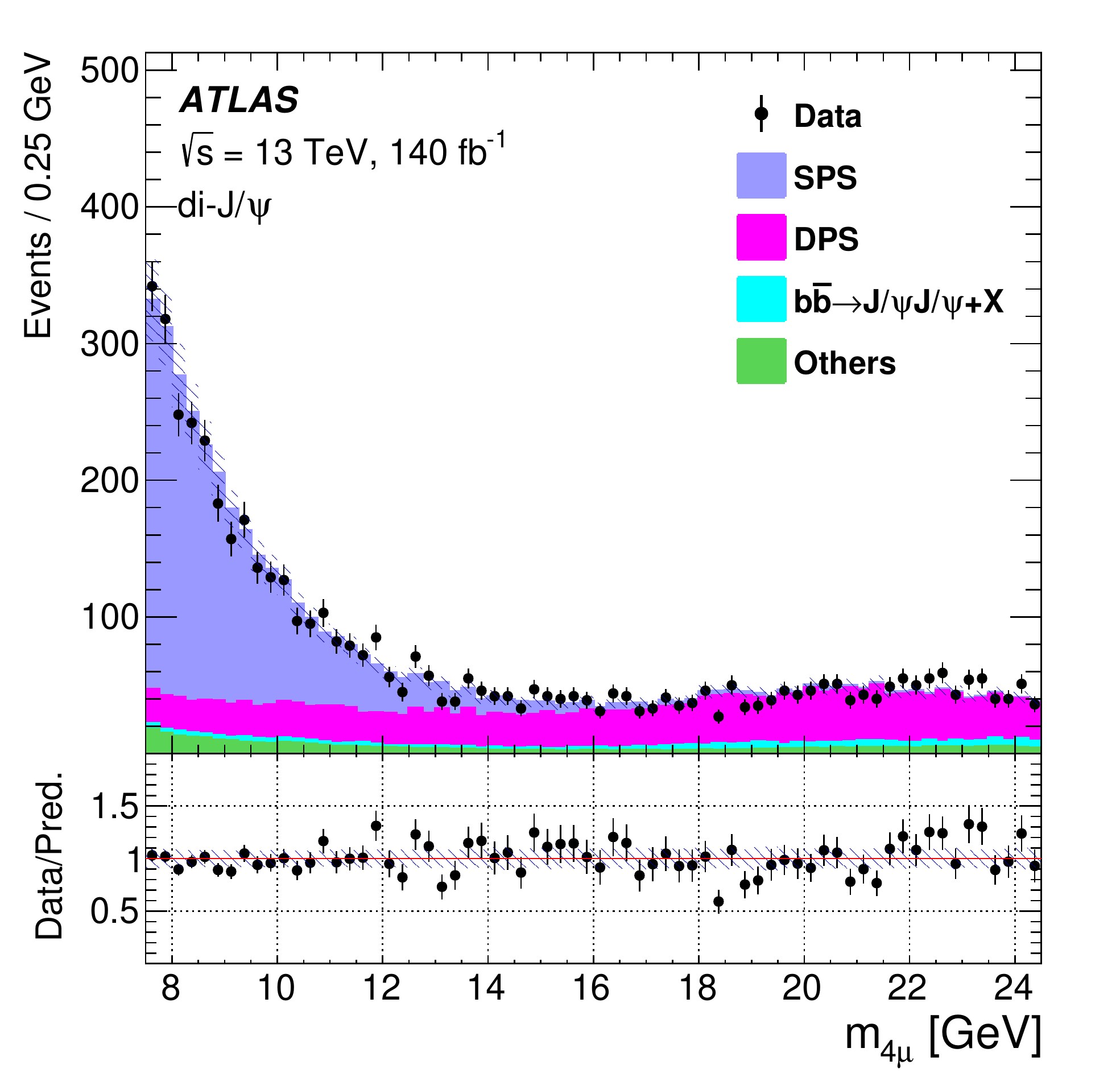}
\put(-45, 70){\textbf{(a)}}
\includegraphics[width=0.33\textwidth]{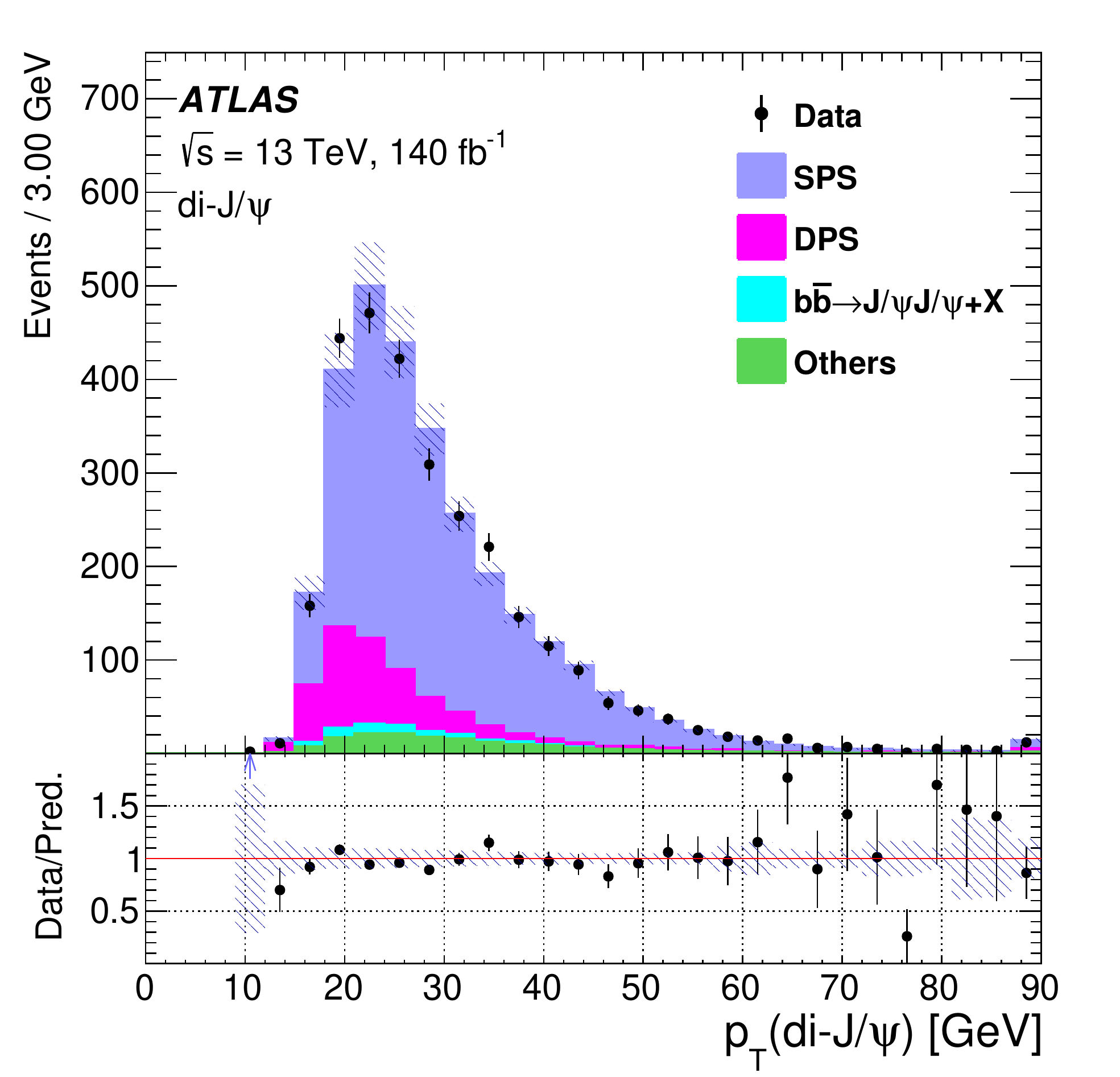}
\put(-45, 70){\textbf{(b)}}
\includegraphics[width=0.33\textwidth]{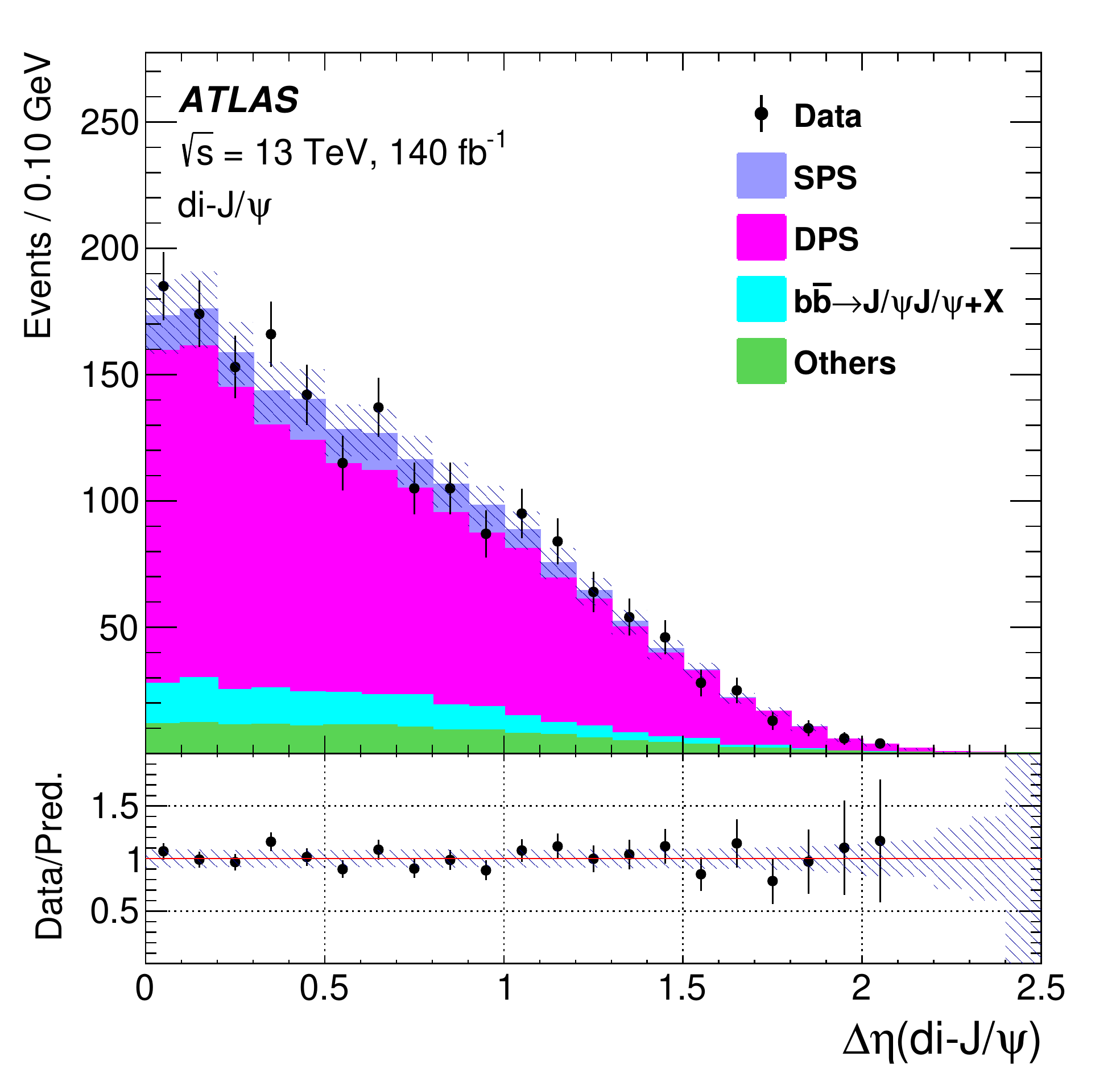}
\put(-45, 70){\textbf{(c)}}
\caption{ The $4\mu$ mass spectrum within $[7.5,~24.5]$ GeV and without the $\Delta R$ requirement~(a), $p_{\text{T}}$ of the di-charmonium in the SPS control region with $7.5~\text{GeV}<m_{4\mu}<12.0~\text{GeV}$~(b), and $\Delta\eta$ between the charmonia in the DPS control region with $14.0~\text{GeV}<m_{4\mu}<24.5~\text{GeV}$~(c), in the di-$J/\psi$ channel. }
\label{fig:control_2jp}
\end{figure}

\begin{figure}[htbp]
\centering
\includegraphics[width=0.33\textwidth]{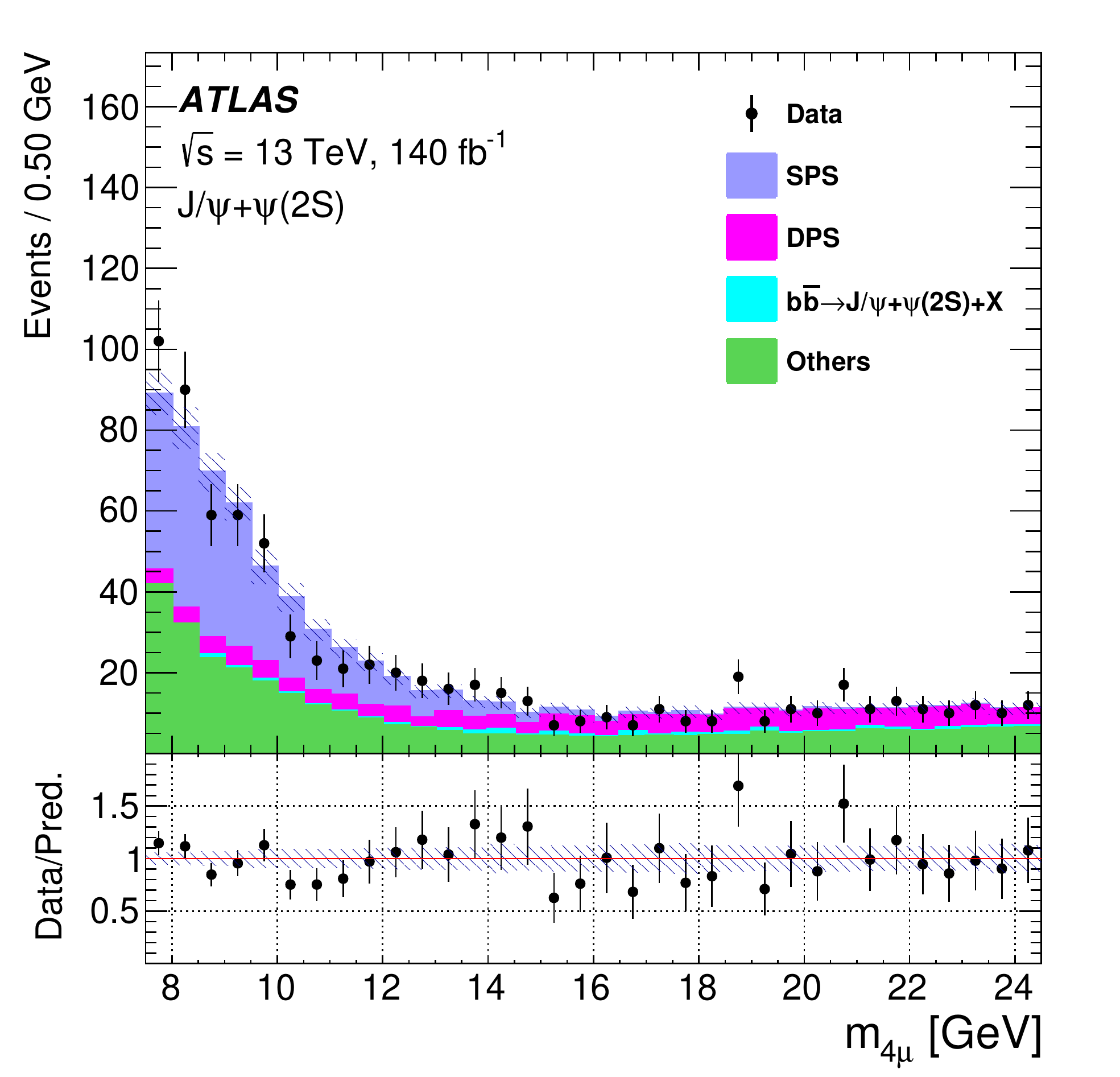}
\put(-45, 70){\textbf{(a)}}
\includegraphics[width=0.33\textwidth]{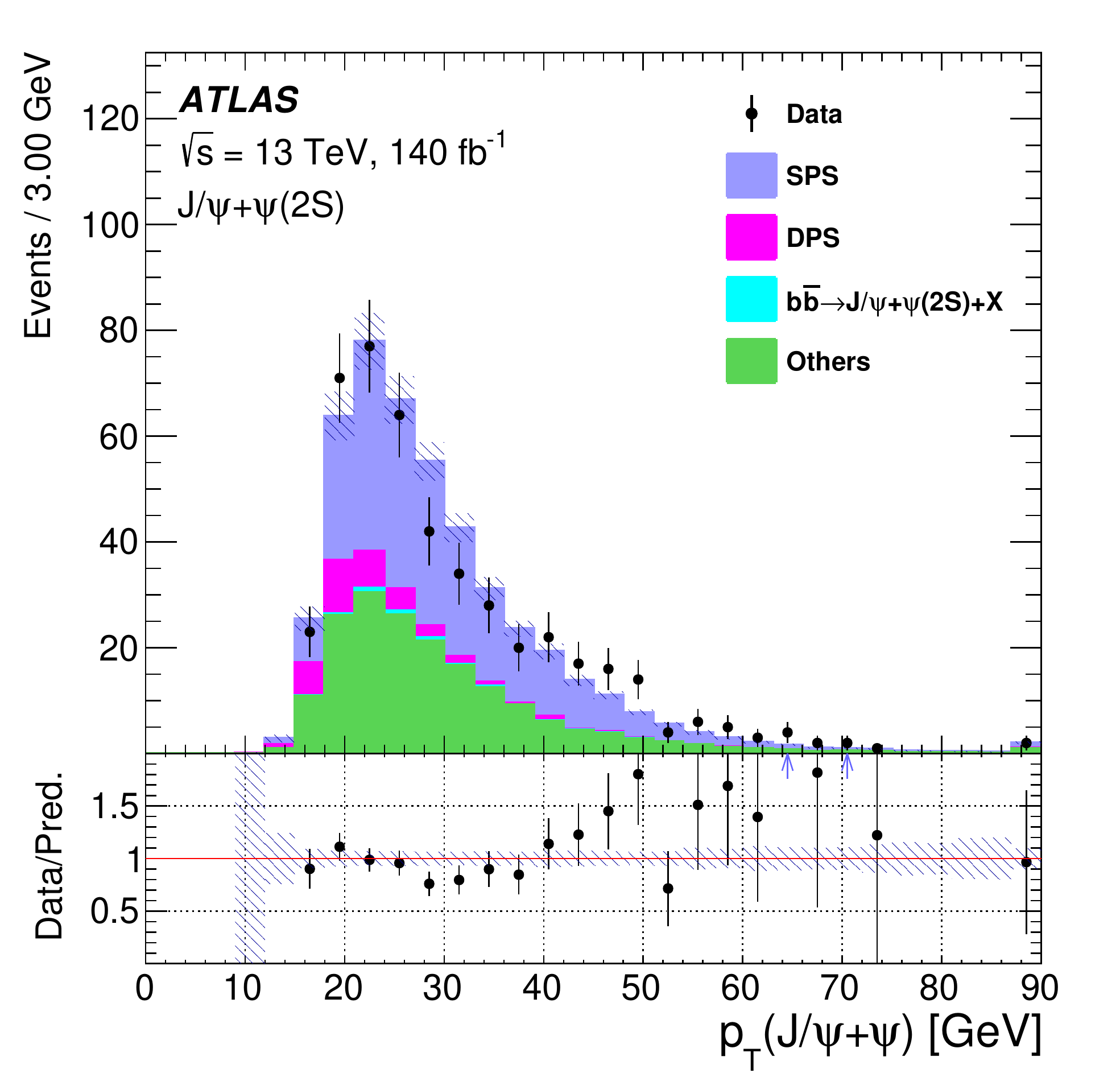}
\put(-45, 70){\textbf{(b)}}
\includegraphics[width=0.33\textwidth]{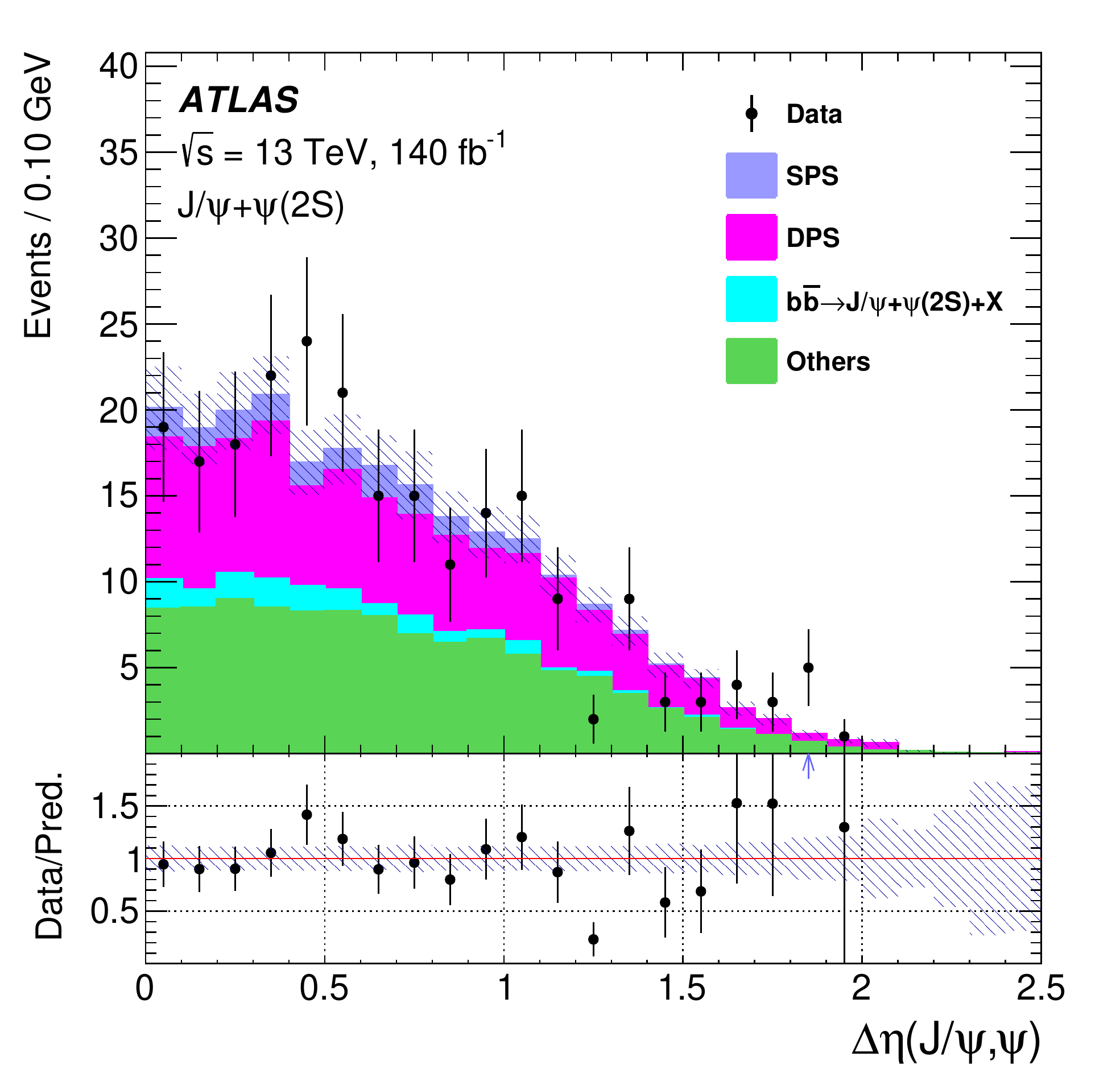}
\put(-45, 70){\textbf{(c)}}
\caption{ The $4\mu$ mass spectrum within $[7.5,~24.5]$ GeV and without the $\Delta R$ requirement~(a), $p_{\text{T}}$ of the di-charmonium in the SPS control region with $7.5~\text{GeV}<m_{4\mu}<12.0~\text{GeV}$~(b), and $\Delta\eta$ between the charmonia in the DPS control region with $14.0~\text{GeV}<m_{4\mu}<24.5~\text{GeV}$~(c), in the $J/\psi$+$\psi$(2S) channel. }
\label{fig:control_jpp}
\end{figure}

The $J/\psi$ and $4\mu$ mass distributions of data and predictions in the signal regions of the two channels before the fits are shown in Figure~\ref{fig:unb_SR}(a,b,c). Similar structures were also observed by CMS \cite{CMS:diJpsi}.
Similar $4\mu$ mass distributions in the control regions are shown in Figure~\ref{fig:unb_CR}. Distributions in the SPS and DPS control regions without the $\Delta R$ requirement between the charmonia are shown in Figure~\ref{fig:control_2jp}-\ref{fig:control_jpp}.
The systematic uncertainties in the fitted masses and widths of the highest resonances in models A and $\alpha$ of the two channels are summarized in Table~\ref{tab:sys_mass_width}.
 
\begin{table}[htbp]
\caption{ Different sources of systematic uncertainty in the mass and natural width (in \MeV) of the third (second) resonance
in model A ($\alpha$) of the di-$J/\psi$ ($J/\psi$+$\psi$(2S)) channel.}
\label{tab:sys_mass_width}
\centering
\renewcommand{\arraystretch}{1.2}
\begin{tabular}{c|cc|cc}
\toprule
Systematic & \multicolumn{2}{c|}{di-$J/\psi$} & \multicolumn{2}{c}{$J/\psi$+$\psi$(2S)} \\
Uncertainties (\MeV) & $m_2$ & $\Gamma_2$ & $m_3$ & $\Gamma_3$ \\ \midrule
Muon calibration & $\pm 6$ & $\pm 7$  & <1 & $\pm 1$ \\
SPS model parameter & $\pm 7$ & $\pm 7$  & \multicolumn{2}{c}{<1}  \\
SPS di-charmonium $p_{\text{T}}$ & $\pm 7$ & $\pm 8$  & \multicolumn{2}{c}{<1} \\
Background MC sample size & $\pm 7$ & $\pm 8$ & $\pm 1$ & <1 \\
Mass resolution & $\pm 4$ & $-3$ & $-1$ & $^{+2}_{-4}$ \\
Fit bias & $-13$ & $+10$  & $^{+9}_{-10}$ & $^{+50}_{-16}$ \\
Shape inconsistency & \multicolumn{2}{c|}{<1} & $\pm 4$ & $\pm 6$ \\
Transfer factor & \multicolumn{2}{c|}{---} & $\pm 5$ & $\pm 23$ \\
Presence of 4th resonance & \multicolumn{2}{c|}{<1} & \multicolumn{2}{c}{---} \\
Feed-down & $^{+4}_{-1}$ & $^{+6}_{-2}$ & \multicolumn{2}{c}{---} \\
Interference of 4th resonance & \multicolumn{2}{c|}{---} & $-32$ & $-11$ \\
P and D-wave BW & $+9$ & $+19$ & <1 & $\pm 1$ \\
$\Delta R$ and muon $p_{\text{T}}$ requirements & $^{+3}_{-2}$ & $^{+6}_{-4}$ & $^{+1}_{-2}$ & $-2$ \\
Lower resonance shape & \multicolumn{2}{c|}{---} & $^{+3}_{-7}$ & $^{+31}_{-34}$ \\
\bottomrule
\end{tabular}
\end{table}

% End of text imported from the .//tex/appendix.tex input file
\clearpage

\printbibliography

\clearpage
% ATLAS Collaboration author list
% Reference date of BPHY-2022-01 is 2023-02-17
% Author list last updated on date 14-OCT-23
% Data extracted on 14-Oct-2023 for paper reference BPHY-2022-01
% at 4:28pm
 
\begin{flushleft}
\hypersetup{urlcolor=black}
{\Large The ATLAS Collaboration}

\bigskip

\AtlasOrcid[0000-0002-6665-4934]{G.~Aad}$^\textrm{\scriptsize 102}$,
\AtlasOrcid[0000-0002-5888-2734]{B.~Abbott}$^\textrm{\scriptsize 120}$,
\AtlasOrcid[0000-0002-1002-1652]{K.~Abeling}$^\textrm{\scriptsize 55}$,
\AtlasOrcid[0000-0001-5763-2760]{N.J.~Abicht}$^\textrm{\scriptsize 49}$,
\AtlasOrcid[0000-0002-8496-9294]{S.H.~Abidi}$^\textrm{\scriptsize 29}$,
\AtlasOrcid[0000-0002-9987-2292]{A.~Aboulhorma}$^\textrm{\scriptsize 35e}$,
\AtlasOrcid[0000-0001-5329-6640]{H.~Abramowicz}$^\textrm{\scriptsize 151}$,
\AtlasOrcid[0000-0002-1599-2896]{H.~Abreu}$^\textrm{\scriptsize 150}$,
\AtlasOrcid[0000-0003-0403-3697]{Y.~Abulaiti}$^\textrm{\scriptsize 117}$,
\AtlasOrcid[0000-0003-0762-7204]{A.C.~Abusleme~Hoffman}$^\textrm{\scriptsize 137a}$
% ...
\bigskip
\\

% ...
$^{29}$Physics Department, Brookhaven National Laboratory, Upton NY; United States of America.\\
% ...
$^{31}$California State University, CA; United States of America.\\
% ...
$^{34}$Department of Physics, Carleton University, Ottawa ON; Canada.\\
$^{35}$$^{(a)}$Facult\'e des Sciences Ain Chock, R\'eseau Universitaire de Physique des Hautes Energies - Universit\'e Hassan II, Casablanca;$^{(b)}$Facult\'{e} des Sciences, Universit\'{e} Ibn-Tofail, K\'{e}nitra;$^{(c)}$Facult\'e des Sciences Semlalia, Universit\'e Cadi Ayyad, LPHEA-Marrakech;$^{(d)}$LPMR, Facult\'e des Sciences, Universit\'e Mohamed Premier, Oujda;$^{(e)}$Facult\'e des sciences, Universit\'e Mohammed V, Rabat;$^{(f)}$Institute of Applied Physics, Mohammed VI Polytechnic University, Ben Guerir; Morocco.\\
% ...
$^{49}$Fakult\"{a}t Physik , Technische Universit{\"a}t Dortmund, Dortmund; Germany.\\
% ...
$^{55}$II. Physikalisches Institut, Georg-August-Universit\"{a}t G\"ottingen, G\"ottingen; Germany.\\
% ...
$^{102}$CPPM, Aix-Marseille Universit\'e, CNRS/IN2P3, Marseille; France.\\
% ...
$^{117}$Department of Physics, New York University, New York NY; United States of America.\\
% ...
$^{120}$Homer L. Dodge Department of Physics and Astronomy, University of Oklahoma, Norman OK; United States of America.\\
% ...
$^{137}$$^{(a)}$Departamento de F\'isica, Pontificia Universidad Cat\'olica de Chile, Santiago;$^{(b)}$Millennium Institute for Subatomic physics at high energy frontier (SAPHIR), Santiago;$^{(c)}$Instituto de Investigaci\'on Multidisciplinario en Ciencia y Tecnolog\'ia, y Departamento de F\'isica, Universidad de La Serena;$^{(d)}$Universidad Andres Bello, Department of Physics, Santiago;$^{(e)}$Instituto de Alta Investigaci\'on, Universidad de Tarapac\'a, Arica;$^{(f)}$Departamento de F\'isica, Universidad T\'ecnica Federico Santa Mar\'ia, Valpara\'iso; Chile.\\
% ...
$^{150}$Department of Physics, Technion, Israel Institute of Technology, Haifa; Israel.\\
$^{151}$Raymond and Beverly Sackler School of Physics and Astronomy, Tel Aviv University, Tel Aviv; Israel.\\
% ...
$^{a}$ Also Affiliated with an institute covered by a cooperation agreement with CERN.\\
$^{b}$ Also at An-Najah National University, Nablus; Palestine.\\
$^{c}$ Also at APC, Universit\'e Paris Cit\'e, CNRS/IN2P3, Paris; France.\\
$^{d}$ Also at Borough of Manhattan Community College, City University of New York, New York NY; United States of America.\\
$^{e}$ Also at Center for High Energy Physics, Peking University; China.\\
$^{f}$ Also at Center for Interdisciplinary Research and Innovation (CIRI-AUTH), Thessaloniki; Greece.\\
% ...

\end{flushleft}

% Created with Glance <Atlas.Glance@cern.ch>

\end{document}